\documentstyle[epsfig]{aipproc}
\def\beq{\begin{equation}}
\def\eeq{\end{equation}}
\def\bea{\begin{eqnarray}}
\def\eea{\end{eqnarray}}
\def\eq#1{{Eq.~(\ref{#1})}}
\def\as{\alpha_{\rm S}}
\begin{document}
\begin{flushright}
 June  22  1998\\
TAUP 2506-98\\
\end{flushright}

\title{EVOLUTION EQUATIONS FOR HIGH PARTON DENSITY QCD}
\author{ Eugene   Levin }
\address{ { \tt leving@post.tau.ac.il}\\
 School of Physics and Astronomy\\
 Raymond and Beverly Sackler Faculty of Exact Science\\
 Tel Aviv University, Tel Aviv, 69978, ISRAEL}
\maketitle

\centerline{}

{\it Talk given at ``Continious Advances in QCD", Minnesota, April 16-19,
1998}

\begin{abstract}
This talk is a brief report on our results in high parton density QCD.
The main goal of this talk is to discuss the new evolution equation which
was obtained and solved in Refs. \cite{AGL} \cite{GLM} and to share with
you the  physics can be recovered by this equation.
\end{abstract}
\section*{Introduction}
In this talk, we  briefly  outline the main ideas, approaches and
results which we developed in high parton density QCD (hdQCD). High
parton density QCD involves a system of partons at short distances
in which the QCD coupling constant $ \alpha_S$ is small, but the
density of partons becomes so large that we cannot apply the usual
methods of perturbative QCD ( pQCD ). In essence, the theoretical
problems here are  non perturbative, but the origin of the non
perturbative effects does not lie in  long distances and large
$ \alpha_S$,  which are present in the confinement region.

First, we need to answer the question:

{ \it 1. Which parton density is high?}

  The quantative estimates of which  density is high, we can obtain from
the
$s$-channel unitarity \cite{GLR} which we can use in two different form:

1.  $\sigma_{tot}( \gamma^* p ) \,\,\leq\,\alpha_{em}\,2\,\pi\,R^2$,
where $R$ is the size of the target, $\alpha_{em}$ is the fine
structure constant ( see  Ref. \cite{GRIB} for explanation why we
have $\alpha_{em}$ ).   This constraint can be rewritten as
\cite{GLR} \cite{MUQI}\cite{MU90}
\beq \label{1}
\kappa\,\,=\,\,\frac{3\,\pi\,\alpha_S x
G(x,Q^2)}{Q^2\,R^2}\,\,\leq\,\,1\,\,.
\eeq
2. $\sigma^{DD}( \gamma^* p )\,\,\leq\,\,\sigma_{tot}( \gamma^* p )$,
where $DD$ stands for diffractive dissociation. This inequality
gives  \eq{1}.

Therefore,

1.  if $\kappa$ is very small ($\kappa \ll 1$ ), we have a
low density QCD in
which
 the parton cascade can be perfectly described by the DGLAP evolution
equations \cite{DGLAP};

2.  if  $ \kappa\,\,\leq\,\,1$, we are in the 
transition region between low and high density QCD . In this region we
can still use pQCD, but have to take into account   the interaction
between
partons inside  the partons cascade;

3. if $\kappa\,\,\geq\,\,1 $, we
reach the region of high parton density QCD. which we are going to discuss
here.

Before doing this we want to make a remark on parton densities in a
nucleus. Taking into account that for nucleus $R_A\,\,=\,\,R_N
\,\times\,A^{\frac{1}{3}}$ and $xG_A(x,Q^2)\,\,=\,\,A\,xG_N(x,Q^2)$, we
can rewrite \eq{1} as
\beq \label{2}
\kappa_A\,\,=\,\,\frac{3\,\pi\,\alpha_S A\, x
G_N(x,Q^2)}{Q^2\,R^2_N\,\,A^{\frac{2}{3}}}\,\,=\,\,A^{\frac{1}{3}}\,\kappa_N
\,\,\leq\,\,1\,\,.
\eeq
Therefore, for the case of an interaction with nucleii, we can reach a
hdQCD region at smaller parton density than  in a nucleon (see Fig.1).

Fig.1 gives
the kinematic plot $(x,Q^2)$ with the line $\kappa = 1 $, which shows that
the hdQCD effect should be  seen  at HERA.

\begin{figure}
\centerline{\psfig{file= 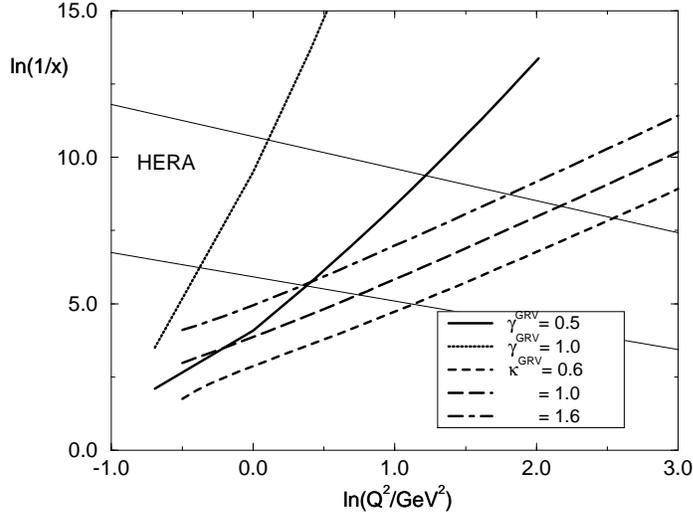,width=100mm}}
\caption{ \em Contours for $< \gamma >$  and $\kappa$   for the GRV'94
gluon density and HERA kinamatic region.}
\label{fig2}
\end{figure}

{\it 2. Two different theoretical approaches}

To understand these two approaches we have to look at the picture of a
high energy interaction in the parton model. In the parton approach the
fast hadron decays into point-like particles ( partons )  long before
( typical time $ \tau \,\propto\,\frac{E}{\mu^2}$ ) the interaction with
the
target. However, during this time $\tau $,   all partons  are in the
coherent state which can be described by means of a wave function. The
interaction of the slowest ($``wee"$) parton with the target completely 
destroys
 the coherence of the partonic wave function.
 \begin{figure}
\centerline{\psfig{file=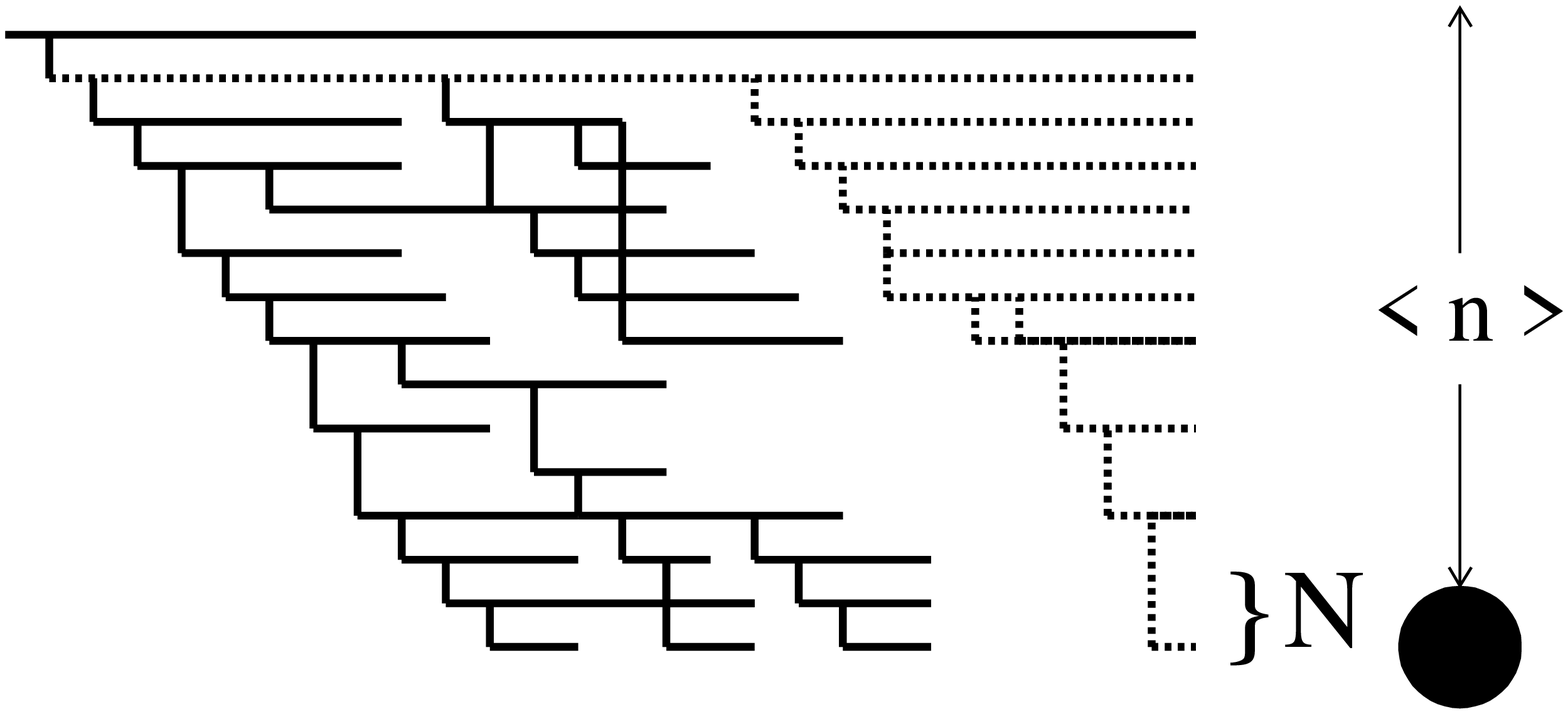,width=100mm}}
\caption{ \em The high energy interaction in the parton model.}
\label{fig3}
\end{figure}

The total cross section of such
an interaction is equal to
\begin{equation}
\label{3}
\sigma_{tot}\,\,=\,\,N\,\times\,\sigma_0
\end{equation}
where
\begin{itemize}
\item  $N$\,\,=
{ \it flux ( }$\mathbf renormalized$ {\bf ?!} { \it ) of ``wee"
partons\,\,;}
\item
$\sigma_0$\,\,=
{ \it the cross section of the
interaction of one
``wee" parton with the target\,\,.}
\end{itemize}
One can see directly from Fig.2 that the number of ``wee" partons is
rather large and it is equal to
\beq \label{4}
N\,\,\propto\,\,e^{ < n>}\,
=\,\frac{1}{x^{\omega_0}}\,\,\,\,with\,\,\,\,
 < n >\,=\,\omega_0
\,\ln(1/x)\,\,;\,\,\omega_0\,=C \alpha_S\,\,.
\eeq
We have to renormalize the flux of ``wee" partons, since the total
cross section is the number of interactions  and if one has  several
``wee" partons with the same momenta they only give  rise to  one
interaction.

If $ N\,\,\approx\,\,1$, we expect that the renormalization of the flux
will be small, and we use an approach with the following typical
ingredients:

$ \bullet$  Parton Approach;
$ \bullet$ Shadowing Corrections;
$ \bullet$ Glauber Approach;
$ \bullet$ Reggeon-like Technique;
$ \bullet$ AGK cutting rules\,.

However, when
$ N\,\,\gg\,\,1$,  we have to change our approach completely from  the
parton cascade to one based on semiclassical field approach, since due to
the uncertainty principle
 $\Delta N \Delta \phi \,\approx\,1$, we can consider the phase as a small
parameter.
Therefore, in this kinematic region our magic words are:

$ \bullet$   Semi-classical gluon fields;
$ \bullet$  Wiezs$\ddot{a}$cker-Williams approximation;
$ \bullet$ Effective Lagrangian for hdQCD;
$ \bullet$  Renormalization Wilson group Approach.

It is clear, that for $N \,\approx\,\,1$ the most natural way is to
approach the hdQCD looking for corrections to the perturbative parton
cascade. In this approach the pQCD evolution has been naturally included,
and it aims to describe the transition region. The key problem
 is to penetrate into the hdQCD region where $\kappa $ is large.
Let us call this approach ``pQCD motivated approach ".

For $N \,\gg\,\,1$, the most natural way of doing is to use the effective
Lagrangian approach,  and remarkable progress has been achieved both in
writing of the explicit form  of this effective Lagrangian,  and in
understanding physics behind it  \cite{EL}. The key problem
for this approach was to find a correspondence  with pQCD. This problem
has been
solved \cite{KOV}.

The {\bf hottest news} of this conference is that these two
approaches
 give the same evolution equation
for the gluon structure function  as reported by A. Kovner \cite{KOVT}.

Fig.3 shows the current situation on the frontier line in the offensive on
hdQCD.

\begin{figure}
\centerline{\psfig{file= 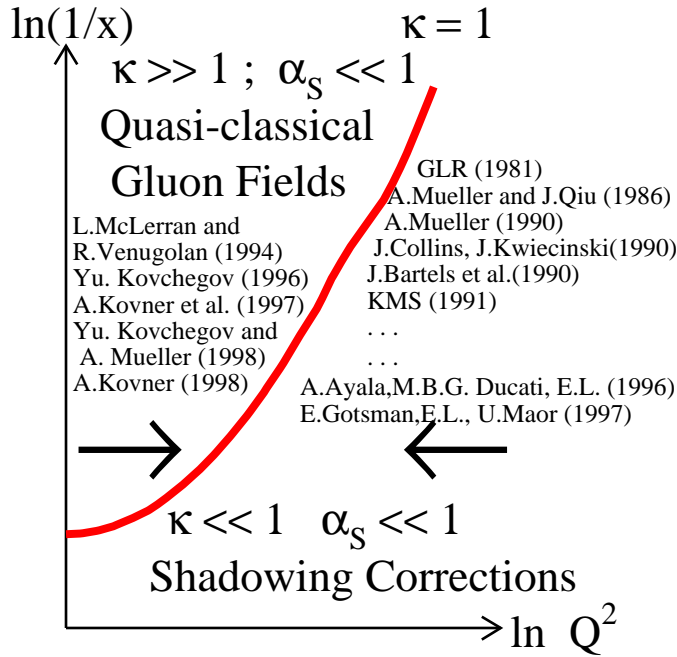,width=100mm}}
\caption{ \em The current situation in the battle for hdQCD.}
\label{fig4}
\end{figure}

{ \it 3. Pluses and Minuses of pQCD motivated approach}

The goal of this talk is to show you how we derive  this evolution
equation
and explain   what physics we can recover using this approach.

Let me list here the main advantages and shortcomings of this approach.
I believe that the honest report on them will lead us to a deeper
understanding of the main theoretical problems in hdQCD.

{  \Huge \bf  +}\,\,{\bf -s\,: }

1. Clear physical meaning   of everything what we are doing\,;

2. Transparent  parameters that we are using for a selection of the Feyman
diagrams\,;

3. Obvious interrelation   with usual DGLAP evolution and the Reggeon
Calculus  ( ``soft" phenomenology )\,;

4. Understandable limitation of the approach which actually means that we
  guarantee the accuracy of our theoretical calculations. In other word,
this approach  has reached a stage of a  normal theory with a regular
procedure of 
how to calculate corrections with respect to small parameters.

5. It has a clear feedback  for such interesting problems as higher
twist contributions, violation of factorization and so on.

{ \Huge \bf  --}\,\,{\bf :}

 There is { \bf no hope} to obtain any formal ( operator ) approach to the
problem and, therefore, to find a formal justification of the resummation
technique. It means, that we can prove everything only in the framework  
of the
Feyman diagrams and in the framework  of the operation rules with them.
However, it means that we ignore, at least partly, non perturbative
effects and that we are not able to formulate a procedure of calculation
of such effects.

Let me illustrate why it is very important an operator proof giving two
examples.

1. In the pQCD motivated approach we heavily use the  so called leading
log
(1/x) approximation in pQCD ( LL(x)A ). The operation rule for LL(x)A is:
\begin{eqnarray} 
& 
xG^{LL(x)A}(x,Q^2)\,\,=\,\,lim|_{x\,\rightarrow\,0} \sum^{\infty}_{n =
1}\,\,
\alpha^n_S\,M_n(x,Q^2) & \label{5}\\
&
=\,\,\sum^{\infty}_{n =
1}\,\,\as^n\,\,lim|_{x\,\rightarrow\,0}
\,M_n(x,Q^2)\,\,=\,\,\sum^{\infty}_{n =
1}\,C_n(Q^2)\,(\,\alpha_S\,\ln(1/x)\,)^n\,\,.
& \nonumber
\end{eqnarray}
We used the gluon structure function as an example. \eq{5} shows
that we changed the order of two operation : summation over $n$ and
limit at $x\,\rightarrow \,0$. Do we have a proof for this? As far
as I know, the answer is {\bf no}. We consider as a strong argument
in favor, the fact that the calculation in the next-to-leading order
in LL(x)A gives a reasonable answer. The real proof  has been done
only recently \cite{KOV} in the effective Lagrangian approach, which
leads to the  LL(x)A of pQCD.

2.  One of the main non perturbative observable which enters any
pQCD motivated calculation is so called the size of the target $R$ ( see
\eq{1} ). We know the meaning of this parameter, namely, this is a
correlation radius between two gluon inside a target at $x\,\approx\,1$.
In the effective Lagrangian approach we will be able to find an operator
expression for this number, and to ask our friends from lattice
QCD to calculate this value. I think this is the only reasonable way of 
how
to take non perturbative QCD into account, at least, at present time.

\section*{New equation}

{\it 4. The GLR equation}

The first attempt to take into account the parton - parton interaction in
the pQCD motivated approach was done  long ago \cite{GLR}. It was based
on the simple idea that there are two processes in a parton cascade (see
 Fig.2) : (i) the probability of the emission of an extra gluon is
proportional to $\alpha_S \,\rho$ where $\rho $ is the density of gluon in
the transverse plane, namely
\beq \label{6}
\rho\,\,=\,\,\frac{xG(x,Q^2)}{\pi\,R^2}\,\,;
\eeq
and (ii) the annihilation   of a gluon,  or in other words a process in 
which  the  probability is proportional to $\rho^2\,\times\,
\sigma_{annihilation}$.
$ \sigma_{annihilation}$ can be estimated as $
\sigma_{annihilation}\,\propto \,\alpha_S\,r^2$, where $r$ is the size of
the parton produced in the annihilation process. For deep inelastic
scattering, $r^2\,\,\propto\,\,\frac{1}{Q^2}$.
Therefore, in the parton cascade we have
\begin{eqnarray}
&
Emission\,\,(\,\,1\,\,\rightarrow\,\,2\,\,)\,:
\,\,\,probability\,\,=\,\,P^{emission}\,\,\propto\,\,\alpha_S\,\,\rho\,\,;
& \label{7}\\
&
Annihilation\,\,(\,\,2\,\,\rightarrow\,\,1\,\,)\,:
\,\,\,probability\,\,=\,\,P^{annihilation}\,\,\propto\,\,\alpha^2_S\,\,
\frac{1}{Q^2}\,\,\rho^2 \,\,.& \label{8}
\end{eqnarray}

At $x\,\sim\,1$ only emission of new partons is essential,  because
$\rho\,\ll\,1$ and this emission is described by the DGLAP
evolution equations. However, at $x \,\rightarrow\,0$ the value of
$\rho$ becomes so large that the annihilation of partons becomes
important,  and so the value of $\rho$ is diminished.  The competition of
these  two processes we can write as an equation for the number of
partons in a phase space cell\footnote{We will argue why we chose this
cell below in section 6.} ( $\Delta y
\,=\,\Delta \,\ln(1/x)\,\Delta \,\ln (Q^2/Q^2_0) $ ):
\beq .\label{9}
\frac{\partial^2 \rho}{\partial \ln (1/x) \partial \ln (Q^2/Q^2_0)}\,\,=
\,\,\frac{\alpha_S \,N_c}{\pi}\,\rho\,\,-\,\,\frac{\alpha^2_S\,\,\tilde
\gamma}{Q^2}\,\,\rho^2\,\,,
\eeq
or in terms of the gluon structure function $xG(x,Q^2)$
\beq \label{10}
\frac{\partial^2 x G(x, Q^2)}{\partial \ln (1/x) \partial \ln
(Q^2/Q^2_0)}\,\,=
\,\,\frac{\alpha_S \,N_c}{\pi}\, x G(x,
Q^2\,\,-\,\,\frac{\alpha^2_S\,\,\tilde
\gamma}{\pi R^2\,\,Q^2}\,\,(\, x G(x, Q^2)\,)^2\,\,.
\eeq
This is the GLR equation which gave the first theoretical  basis
for the  consideration  of hdQCD.  This equation describes the
transition region at very large values of $Q^2$, but  a glance  at Fig.1
shows that we need a tool to  penetrate the kinematic
region of moderate and even small $Q^2$.

{\it  5. Glauber - Mueller Approach}

We found that it is very instructive to start with the Glauber
approach to SC.The idea of how to write the Glauber formula in QCD was
originally formulated in two papers Ref.\cite{LR87} and Ref.
\cite{MU90}. However, the key paper for our problem is the second
paper of A. Mueller, who considered the Glauber approach for the
gluon structure function. The key observation is that the  fraction of
energy,  and the transverse coordinates of the fast  partons can be
considered as frozen,  during the high energy interaction with the
target \cite{LR87} \cite{MU90}. Therefore, the cross section of the
absorption of gluon($G^*$) with virtuality $Q^2$ and Bjorken $x$
can be written in the form:
\beq \label{11}
\sigma^N_{tot}(\,G^*\,)\,\,=\,\,
\eeq
$$
\int^1_0 d z \,\,\int \,\frac{d^2 r_t}{2 \pi}\,\,
\int\frac{d^2 b_t}{2 \pi}
\Psi^{G^*}_{\perp} (Q^2, r_t,x,z)\,\,
 \sigma_N (x,r^2_t)\,\,\,\,[{\Psi^{G^*}_{\perp}} (Q^2, r_t,x,z)]^* \, ,
$$
where $z$ is the fraction of energy
which is carried by the gluon, $\Psi^{G^*}_{\perp}$ is the wave
function of the transverse polarized gluon,  and $\sigma_N (x,r^2_t)$ is
the
cross section of the interaction of  the $GG$- pair with transverse
separation
$r_{t}$ with the nucleus.
Mueller showed that \eq{11} can be reduced to
\beq \label{MF}
x G^{MF}(x,Q^2) = \frac{4}{\pi^2} \int^{1}_{x} \frac{d x'}{x'}
\int_{\frac{1}{Q^2}}^{\infty} \frac{d^2 r_t}{\pi r_{t}^{4}}
\int_{0}^{\infty} \frac{d^2 b_t}{\pi}  2
\left\{ 1 - e^{- \frac{1}{2}\,\sigma_{N}^{GG} ( x^{\prime},r^2_t )
S(b^2_t) }
\right\}
\eeq
with
$$
\sigma_{N}^{GG}\,\,=\,\,\frac{4 \,\pi^2\,\alpha_S }{3}\,\,r^2_t\,x
G^{DGLAP}(x,\frac{1}{r^2_t})\,\,
$$
and profile function $S(b_t)$ chosen in the Gaussian form:
$
S(b^2_t)\,\,=\,\,\frac{1}{\pi R^2}\,\,e^{-\,\frac{b^2_t}{R^2}}\,\,.
$

Obviously, the Mueller formula has a defect, namely, only the fastest
partons ($GG$ pairs) interact with the target.
This assumption is an artifact of the Glauber approach, which looks
strange in
 the parton picture of the interaction. Indeed, in the parton model
we   expect that all partons not only the fastest ones should
interact
 with the target.  At first sight we can solve this problem by  iteration
\eq{MF}( see Ref. \cite{AGL} ). It means that the first iteration will
take into account that not only the fastest parton, but the next one will
interact with the target, and so on.

{\it 6. Why equation?}

We would like to suggest a new approach based on the  evolution
equation
 to sum all SC. However, we  want to argue first,  why an equation is
 better than any iteration procedure. To illustrate this point of view, 
let
us
differentiate the Mueller formula with respect to $y\,=\,\ln(1/x)$ and $
\xi\,=\,\ln Q^2$. It is easy to see that this derivative is equal to
\beq \label{DER}
\frac{\partial^2 x G(x, Q^2)}{\partial y\, \partial \xi}\,\,=\,\,
\frac{4}{\pi^2}\,\int d b^2_t
\,\,\{\,\,1\,\,-\,\,e^{-\,\frac{1}{2}\,\sigma(x,
r^2_{\perp}\,=\,\frac{1}{Q^2})\,S(b^2_t)}\,\,\}\,\,.
\eeq
The advantages  of \eq{DER} are
\begin{itemize}
\item Everything enters at small
distances\,\,;

\item  Everything is under theoretical control\,\,;

\item  Everything that is not known  ( mostly  non perturbative )
is hidden in the initial and boundary condition.
\end{itemize}

Of course, we cannot get rid of our problems changing the procedure
of solution. Indeed, the non-perturbative effects coming from the
large distances are still important,  but they are absorbed in the
boundary and initial conditions of  the equation. Therefore, an
equation is a good ( correct ) way to separate  what we know (
small distance contribution) from what we don't ( large distance
contribution).

{\it 7. Equation}

We suggest the following way to take into account the interaction of all
partons in a parton cascade with the target.
Let us differentiate the $b_t$-integrated Mueller
formula of \eq{MF} in $y \, = \,\ln (1/x)$ and $ \xi = \ln(Q^2/Q^2_0)$.
 It gives
\beq \label{EQ}
\frac{\partial^2 x G ( y,\xi)}{\partial y \partial \xi}\,\,=\,\,
\frac{2 \,Q^2 R^2 }{ \pi^2}\,\,\,\left\{\,\, C \,\,+ \,\,\ln(\kappa_{G} (
x',
 Q^2 )) \,\,+\,\,
E_1 (\kappa_{G} ( x', Q^2 ))  \right\} \,\,\,,
\eeq
where $\kappa_G (x,Q^2)$ is given by
\beq \label{KAPPA}
\kappa ( x,Q^2) \,\,=\,\,
\frac{N_c \alpha_S \pi }{2 Q^2 R^2 }\,x G(x,Q^2)\,\,.
\eeq
Therefore, we consider $\kappa_G$ on the l.h.s. of \eq{EQ} as the
observable which is written through the solution of \eq{EQ}.

\eq{EQ} can be rewritten in the form ( for fixed $\as$
)
\beq \label{EQKA}
\frac{\partial^2 \kappa_G( y,\xi)}{\partial y \partial \xi}\,\,+\,\,
\frac{\partial \kappa_G(y, \xi)}{\partial y}\,\,=\,\,
\frac{ N_c\, \as}{\pi}\,\, \left\{\,\, C \,\,+ \,\,\ln(\kappa_{G})
\,\,+\,\,
E_1 (\kappa_{G})  \right\}
\,\,\equiv\,\,F(\kappa_G)\,\,.
 \eeq
This is the equation which we propose.

{\it 8. Nice properties of the equation:}

This equation has the following desirable properties:

\begin{enumerate}
\item   It  sums all contributions of the order $
(\,\alpha_S\,y\,\xi\,)^n$, 
absorbing them in $x G (y,\xi)$, as well as all contributions of the order
of $\kappa^n$.
Therefore, this equation solves the old problem, formulated in
Ref.\cite{GLR},
and
 for $N_c\,\rightarrow \,\infty $ \eq{EQKA} gives the complete
solution to our problem, summing all SC\,\,;

\item The solution of this equation matches with the solution of the DGLAP
 evolution equation in the DLA of perturbative QCD at $\kappa\,\rightarrow
\,0$\,\,;

\item   At small values of $\kappa$ ( $\kappa\,<\,1$ )
 \eq{EQKA} gives the GLR equation. Indeed,
for small $\kappa$ we can expand the r.h.s of \eq{EQKA} keeping
only the
 second term. Rewriting the equation through the gluon structure function
 we obtain
  the GLR equation \cite{GLR} with
 the coefficient in front of the second term as 
 calculated by Mueller and Qiu \cite{MUQI}\,\,;

\item The first iteration of this equation gives the Mueller formula ( see
 Ref. \cite{MU90})\,\,;

\item In general, everything that we know about SC is included in \eq{EQKA}\,\,;

\item As has been mentioned, \eq{EQKA} was reproduced  
by the effective Lagrangian approach
(see A. Kovner talk at this conference \cite{KOVT}\,\,.

\end{enumerate}

{\it 9.  The theory status of the equation }

Here, we list all small ( large ) parameters that have been used to
obtain \eq{EQKA}:
\begin{itemize}
\item \begin{eqnarray}
&
   \alpha_S\,\ln (1/x)\,\ln(Q^2/Q^2_0)\,\,\approx\,\,1\,\,;
&\label{12}\\ &
\alpha_S\,\ln (1/x)\,\,<\,\,1\,\,;\,\,\,\,\,\,\,\alpha_S\,\ln (Q^2/Q^2_0)\,\,<\,\,1\,\,;
& \label{13} \\ &
\alpha_S\,\,\,\ll\,\,\,1\,\,.& \label{14}
\end{eqnarray}
These equations (\eq{12} - \eq{14} )  give  the conditions when we
can absorb all $ (\,\alpha_S\,\ln (1/x)\,\ln (Q^2/Q^2_0)\,)^n$ -
terms in the gluon structure function both in \eq{EQKA} and in
\eq{MF}.

\item 
\beq \label{15}
\kappa\,\,\leq\,\,\frac{1}{\alpha_S}\,\,.
\eeq
This is the main limitation of our resummation which stems from the
contribution of so called enhanced diagrams ( see Ref. \cite{AGL}
for detail).

\item \beq \label{16}
N_c\,\,\gg\,\,1\,\,.
\eeq
where $N_c$ is the number of colours.  We follow the Veneziano
principle \cite{VEN} of large $N_c$ resummation: we select the
diagrams in leading order with respect to $(\,\alpha_s \,N_c\,)^n$
in each topologic configuration (see Ref. \cite{AGL} and references
therein for a detail discussion of this point ).

\end{itemize}
\section*{Solution}

{\it 10. Asymptotic solution}

A first observation is the fact that \eq{EQKA} has a solution which
depends only
 on $y$. Indeed, one can check that $\kappa \,=\,\kappa_{asymp}(y)$ is
the solution of the following equation
\beq \label{ASYM}
\frac{d \kappa_{asymp}}{d y}\,\,=\,\,F(\,\kappa_{asymp}\,)\,\,.
\eeq
The solution to the above equation is
\beq \label{ASSOL}
\int^{\kappa_{asymp}(y)}_{\kappa_{asymp}(y=y_0)}\,\,
\frac{ d \kappa'}{F(\kappa')}
\,\,=\,\,y - y_0\,\,.
\eeq
one can see that \eq{ASSOL} leads to
\bea
&
\kappa_{asymp}(y)\,\,\rightarrow\,\,\as\, y\,\ln (\as y )
\,\,\,\,at\,\,\as
y\,\,\gg\,\,1\,\,; 
& \label{ASYMSOL}\\
&
or\,\,\,\,\,
x G(x,Q^2)\,\,\propto\,\,Q^2\,R^2 \ln(1/x)\,\ln
(\as\ln(1/x))\,\,\,at\,\,x\,\,\longrightarrow\,\,0\,\,&\nonumber
\eea
We claim this  solution is the asymptotic solution of  \eq{EQKA}. To
prove this we have to consider the stability of the asymptotic
solution. It means, that we solve our general equation looking for
the solution of  the form
\beq \label{STAB}
\kappa\,\,=\,\,\kappa_{asymp}(y)\,\,+\,\,\Delta \kappa(y,\xi - \xi_0)\,\,,
 \eeq
where $\Delta \kappa$ is small ($ \Delta
\kappa\,\ll\,\kappa_{asymp}$) but an arbitrary function at $\xi
\,=\,\xi_0$. We have to prove that \eq{STAB}
 will not lead to big $\Delta \kappa$ ($\Delta
\kappa\,\gg\,\kappa_{asymp}$) at large $\xi$.
The following linear equation can be written for $\Delta \kappa$
\beq \label{109}
\frac{\partial^2 \Delta \kappa( y, \xi  )}{\partial y\,\partial \xi}\,\,+\,\,
\frac{\partial \Delta \kappa( y, \xi)}{\partial y}\,\,=\,\,\frac{d F(\kappa)}{
d \kappa}\,|_{\kappa = \kappa_{asymp}(y)}\,\,\Delta \kappa (y, \xi)\,\,.
\eeq
In Ref.\cite{AGL} it  was proven, that the solution of \eq{STAB} is
much smaller
 than $\kappa$, since $\frac{d
F(\kappa)}{d \kappa}\,\,\rightarrow\,\,0$ at large $y$.

Therefore, the asymptotic solution has a chance to be the solution of
\eq{EQKA}  in the
region of very small $x$. To prove this,
we need to solve the equation
in the wide kinematic region starting with the initial condition.
 We managed to do this only in the semiclassical approach.

{\it 11. The semiclassical approach}

In the semiclassical approach the solution of \eq{EQKA}  is
supposed to be of  the form
\beq \label{S1}
\kappa = e^S \, ,
\eeq
where $S$ is a function with partial derivatives:
$\frac{\partial S}{
\partial y} = \omega $ and $\frac{\partial S}{\partial \xi} = \gamma $
which are smooth functions of $y$ and $\xi$.

 For an equation in the form
$
F(\xi, y, S, \gamma , \omega ) = 0 \, ,
$
we can introduce the set of characteristic lines $ (\xi(y), S(y), \omega 
(y),
\gamma (y) ) $, which satisfy a set of well defined equations
(see, for example, Refs. \cite{Collins90}, \cite{Bartels91} for the
method). In Ref.  \cite{AGL} we developed a detailed calculation of
the solution. The set of equations is:
\beq \label{S3}
\frac{d \xi}{d y}\,\, = \,\,\frac{\Phi(S)}{(\gamma +1)^2}  \,\, ;
\,\,\,\,
\frac{d S}{d y} \,\,= \,\,  \frac{2 \gamma + 1}{(\gamma +1)^2}\Phi(S) \,\,
;\,\,\,
\frac{d \gamma}{d y} = \Phi'{S} \frac{\gamma}{\gamma +1} \, \,;
\eeq
where $\Phi'_{S}\,=\, \frac{\partial \Phi(S}{\partial S}$. The
initial conditions for this set of equations
\beq
 S_0 \,\, = \,\, ln \kappa_{in} (y_0, \xi_0)\,\,;\,\,\,\,
\gamma_0 \,\, = \,\, \left. \frac{\partial \ln \kappa_{in} (y_0 , \xi )}
{\partial \xi} \right|_{\xi = \xi_0} \,\, .
\label{S7}
\eeq
The main properties of these equations have been considered in
Ref.\cite{AGL} analytically. Here, however,
 we restrict ourselves mostly to the numerical   solution
of these equations.

\begin{figure}[hptb]
\begin{center}
\begin{tabular}{ c  c}
\psfig{file=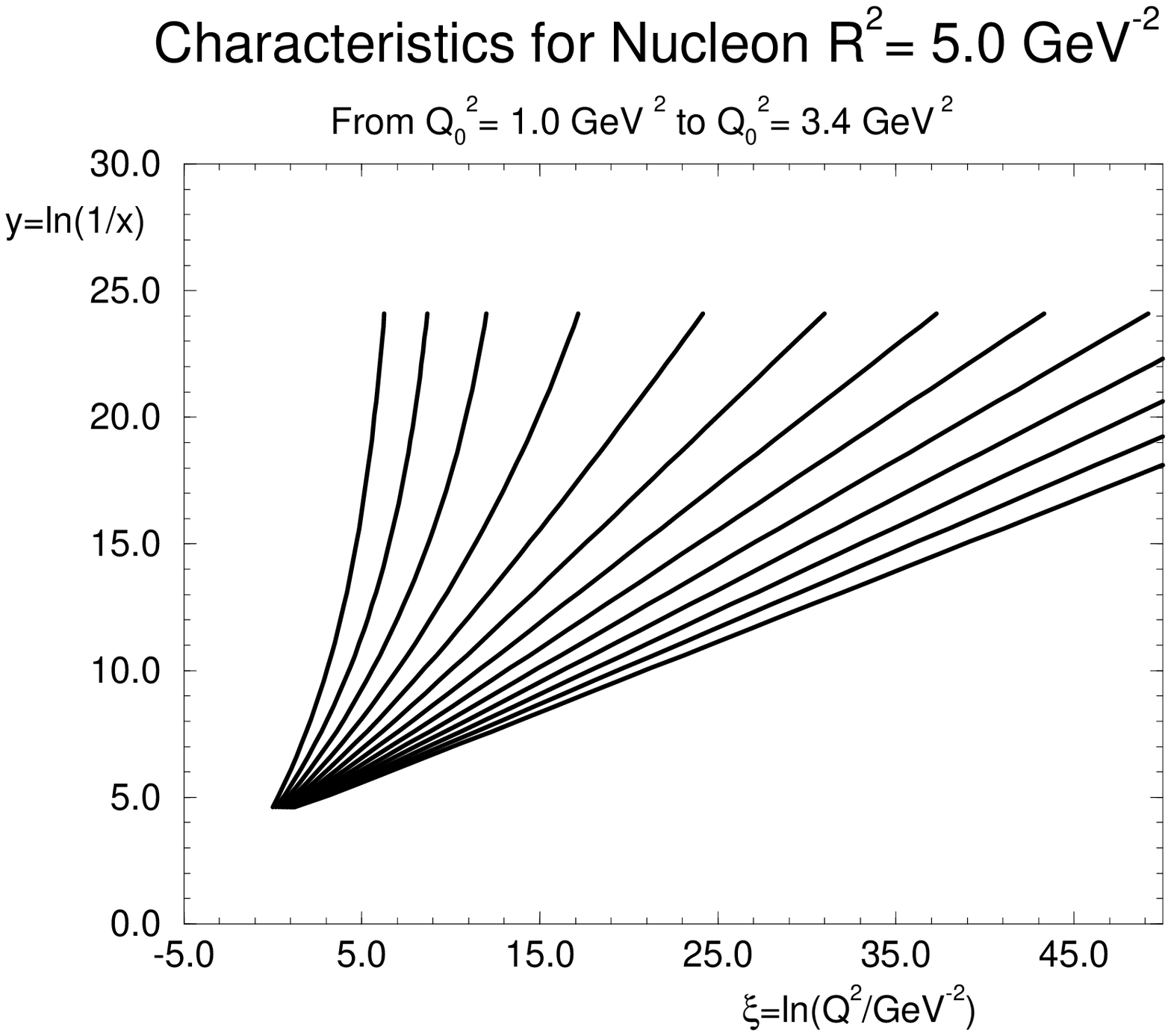,width=65mm} &
\psfig{file=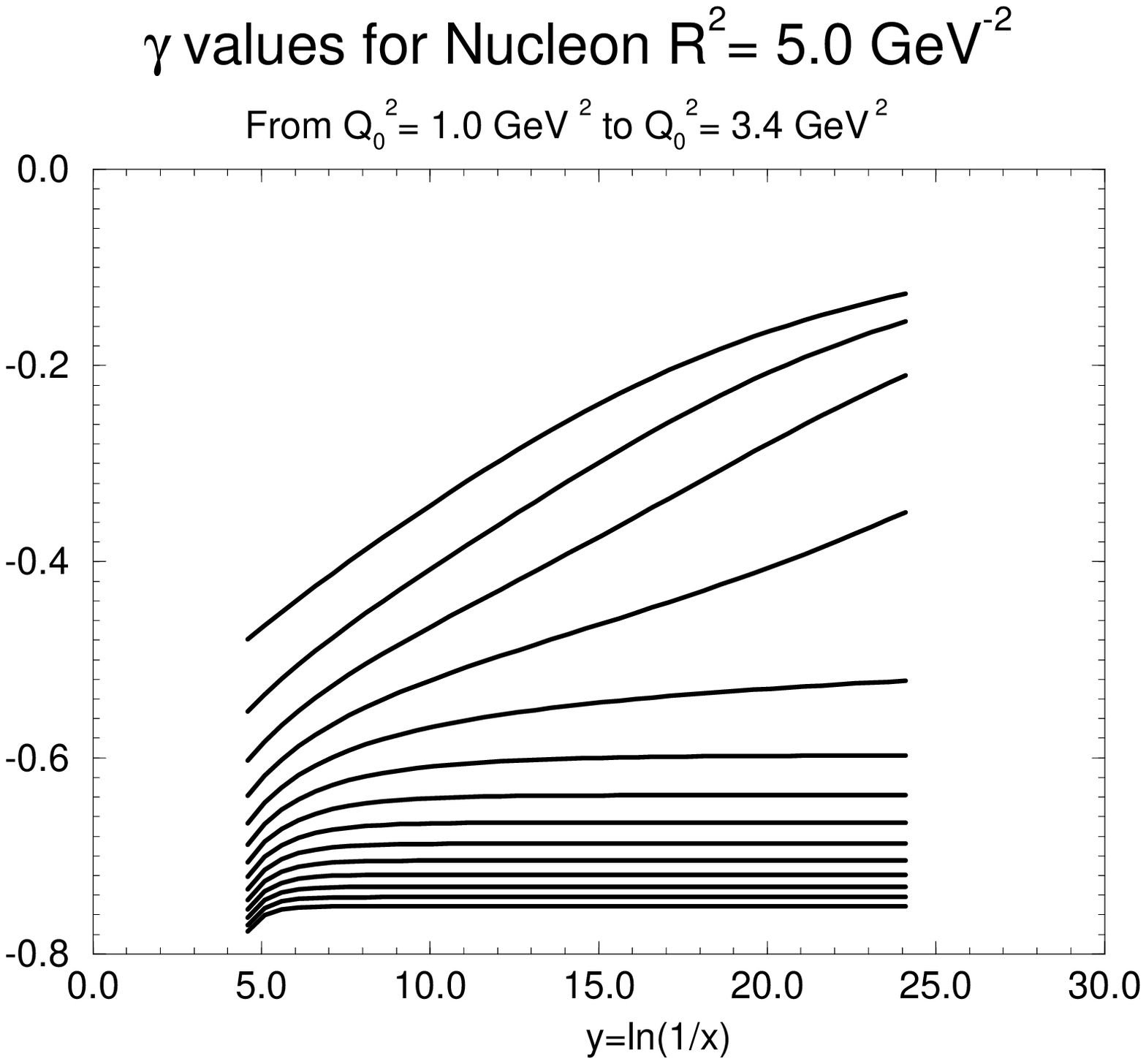,width=65mm}\\
\end{tabular}
\end{center}
\caption{ \em The trajectories (a) and the $\gamma$ values (b) for the solution of
the generalized
 evolution equation for nucleon with $R^2 = \, 5 \, GeV^{-2}$.}
\label{scn}
\end{figure}

Fig. \ref{scn} shows the behaviour of trajectories and $ \gamma$
value for the solution. One can see that the trajectories can be
divided in two groups: (i) first with $\gamma < - \frac{1}{2}$,  and
all of them approach the trajectories of the DGLAP evolution
equation at large $Q^2$; and (ii) second one with $\gamma\,\,>\,\,-
\,\frac{1}{2}$ and the solution of \eq{EQKA} on these trajectories
tends to the asymptotic solution at $y\,\,\gg\,\,1$.

We consider this figure as a nice illustration of our statement
that the high energy (low $x$)  limit of the general solution to
\eq{EQKA} is the asymptotic solution.

{\it 12. Comparison with other approaches}

\begin{figure}
\centerline{\psfig{file= 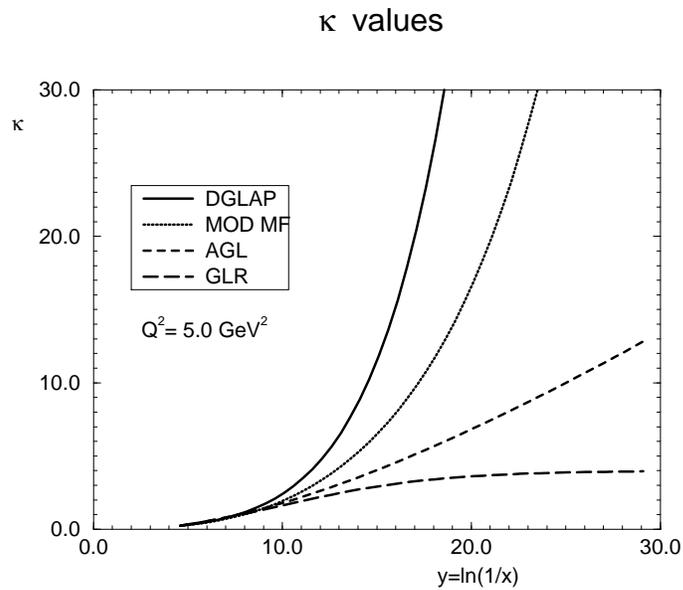,width=100mm}}
\caption{ \em The current situation in the battle for hdQCD.}
\label{scc}
\end{figure}

In Fig.\ref{scc} we plot our calculation for the DGLAP evolution
equation, for the Glauber - Mueller approach, for the GLR equation
and for the new evolution equation. We can conclude, that:
\begin{enumerate}
\item SC even in the Glauber - Mueller approach are essential
for the gluon structure function\,\,;

\item The Glauber - Mueller approach considerably underestimates
 the value of SC\,\,;

\item The GLR equation leads to stronger SC than the solution to \eq{EQKA}\,\,;

\item The new evolution equation does not reproduce the saturation of the gluon density
in the region of small $x$ which the GLR equation leads to\,\,.
\end{enumerate}

\section*{Summary}

{\it 13. Where are SC?}

We  firmly believe that the new evolution equation gives  the 
correct way of evaluation of the value of SC. However, the
difficult question  arises: why the SC have not been seen at HERA?
Our answer is:

1. The value of SC is rather large  but only in the gluon structure
function while their contribution to $F_2$ is rather small
\cite{AGL},  and cannot be seen on the background of the experimental
errors\,\,;

2. The theoretical  determination of the gluon
structure function is not very presize  and we evaluate the errors  as
50\%
\cite{LERIHC} and the SC in $xG$ is hidden in such  large
errors\,\,;

3. The statement that SC have not been seen is also not quite
correct. Our estimates \cite{GLMSLOPEF2} show that the contribution
of SC is rather large in the $F_2$ slope and incorporating  the SC one can
describe the recent experimental data \cite{DATACOLD} on
$\frac{\partial F_2(x,Q^2)}{\partial \ln (Q^2/Q^2_0)}$ (see
Ref.\cite{GLMSLOPEF2} for detail ).

{\it 14. Last words }

I hope that I gave you a brief but to some extent full review of
our approach. Obviously, you can obtain more information from two
large papers of Ref. \cite{AGL}. Deeply in my heart,  I firmly
believe that there will be a bright  future for the effective Lagrangian
approach in  which one 
combines the physics of hdQCD, with the formal methods of the
quantum field theory, and gives  a way to incorporate the lattice
calculation for the non-perturbative observables. However, the
parameter or better to say a newscale  $Q^2_0(x)$  which is the
solution of the equation $ \kappa(x,Q^2_0(x))\,\,=\,\,1 $ only can be
found  in the pQCD motivated approach, since the effective Lagrangian was
 only derived  and justified assuming this new scale.
The fact that the
correct equation was first found in this approach is also not
accidental . since the pQCD motivated approach has a great
advantage : clear understanding of physics of hdQCD.

 \end{document}